\documentstyle[floats,prl,aps,twocolumn,graphicx]{revtex}
\def\MSA{$\mu$-\textsc{squid}~}
\def\MSB{$\mu$-\textsc{squid}}
\begin{document}
\draft
\title{Observation of Persistent Currents in Mesoscopic Connected Rings} 
\author{W. Rabaud$^{1}$, L. 
Saminadayar$^{1}$, D. Mailly$^{2}$, K. 
Hasselbach$^{1}$, A. Beno\^\i t$^{1}$ and B. Etienne$^{2}$} \address{$^{1}$Centre
de Recherches sur les Tr\`es Basses Temp\'eratures, B.P. 166 X, 38042
Grenoble Cedex 9, France\\ $^{2}$Laboratoire de Microstructures et 
Micro\'electronique, B.P. 107, 92225 Bagneux Cedex, France}
\date{\today}
\maketitle

\begin{abstract}
We report measurements of the low temperature magnetic response of a line of $16$ 
GaAs/GaAlAs \emph{connected} mesoscopic rings whose total length is much 
larger than $l_{\phi}$. Using an 
on-chip micro-\textsc{squid} technology, we have measured a periodic response, 
with period $h/e$, corresponding to persistent currents in the rings 
of typical amplitude $0.40\pm 0.08\, nA$ per ring. Direct comparison 
with measurements on the same rings but \emph{isolated} is presented.
\end{abstract}

\pacs{PACS numbers: 73.23.-b, 73.20.Dx, 72.20.My}

In a mesoscopic metallic sample, quantum coherence of the electronic wave 
functions can affect drastically the equilibrium properties of the 
system. In the case of a metallic ring in a magnetic field, the new boundary 
conditions imposed by the magnetic flux\cite{Byers} lead the wave 
functions and therefore all the 
thermodynamic properties of the system to be periodic with flux, with 
periodicity $\Phi_{0}=h/e$, the flux quantum. One of the most striking 
consequences of this, first 
pointed out by B\"{u}ttiker, Imry and Landauer\cite{Butt1} for the 1D case, is that a 
mesoscopic normal-metal ring pierced by a magnetic flux carries a persistent non-dissipative current: this is a consequence of the periodicity of the free 
energy ${\mathcal{F}}(\Phi)$ with flux, implying the existence of an equilibrium
current $I(\Phi)=-\partial{\mathcal{F}}(\Phi)/\partial{\Phi}$. 
Subsequently, much 
theoretical effort has been devoted to the description of a realistic 
3D disordered ring\cite{Mont1,Ambega,Opp1,Cheung,Mont5}. Both 
the sign and the amplitude of this current depend on the number of 
electrons in the ring and on the microscopic disorder configuration: 
thus this current, like other mesoscopic phenomena such as Aharonov-Bohm 
conductance oscillations\cite{Washburn}, is sample specific. However, 
the order of magnitude of 
the current for a single, isolated ring can be characterized by its 
typical value $I_{typ}=\sqrt{\langle 
I^2\rangle}$, $\langle\,\rangle$ denoting the average over disorder configurations. 
It is given\cite{Cheung,Mont5} by $I_{typ} = 1.56\; I_{0}\; l_{e}/L$. In 
this formula, $I_{0}=ev_{F}/L$ 
where $v_{F}$ is the Fermi velocity and $L$ the perimeter of the 
ring, and $l_{e}$ is the elastic mean free path: the current varies 
as the inverse of the diffusion time $\tau_{D}=L^{2}/D$ where 
$D=v_{F}\, l_{e}$ is the diffusion constant. When measuring an ensemble of rings, 
the typical current per ring decreases as $1/\!\sqrt{N_{R}}$, where $N_{R}$ is 
the number of rings. At finite temperature\cite{Cheung,Mont5,Land1}, mixing of the energy levels reduces the current 
on the scale of the Thouless energy $E_{c}$, the energy scale for 
energy correlations. Further reduction arises when temperature reduces 
the phase coherence length $l_{\phi}$ down to values lower than $L$.

For a long time, persistent currents were believed to be a 
specific property of isolated systems\cite{Butt1}. However, recent 
theoretical predictions suggest that persistent currents should exist 
even in connected rings. Using a semi-classical model in the 
diffusive limit, ref.\cite{Mont3} calculated persistent currents in various 
networks of connected rings, showing that the amplitude of the 
persistent currents should be reduced only weakly as compared to its 
value in the same network of isolated rings, whereas all the other 
properties, like temperature dependance, should be similar. The reduction factor 
$r=I_{connected}/I_{isolated}$ depends on the geometry of the network. 
For a line of connected rings separated by arms longer than $l_{\phi}$, 
a simple fraction $r=2/3$ was predicted\cite{Mont3}; for the 
geometry considered here, it is expected\cite{Mont4} that $r\approx 0.58$. Moreover, this 
result is independent of the total size of the system, even much larger 
than $l_{\phi}$. This suggests that persistent currents 
could be observed in a macroscopic system. However, no experimental evidence of 
persistent currents in such a connected geometry has been reported up to 
now.

A couple of key experiments have confirmed the existence of persistent 
currents in isolated systems, either an 
ensemble of isolated rings\cite{Levy,Reulet,Mohanty}, or a single isolated 
ring\cite{Chandra,Mailly}. However, the amplitude of the currents 
found in ref.\cite{Levy} and ref.\cite{Chandra}, much larger than 
expected, is still not 
understood. In this context, the need for 
new experiments to clarify these results and provide new experimental 
facts is very important. Moreover, the existence of persistent 
currents in a network of connected rings larger than $l_{\phi}$ remains 
to be experimentally demonstrated.

In this Letter, we report on measurement of the low temperature magnetic response of a 
line of $16$ GaAs/GaAlAs \emph{connected} rings. The experiment was 
performed at the base temperature of the dilution fridge in order to maximize 
the signal. The sample was designed so that its total size was much larger than $l_{\phi}$, while 
the perimeter of each ring was smaller than $l_{\phi}$. We developed a 
new experimental setup based on a multiloop \MSA gradiometer, and observed 
a periodic response of the magnetization with period $\Phi_{0}=h/e$, corresponding to persistent currents in 
the rings of amplitude $0.40 \pm 0.08\, nA$ per ring. 
Using gates, we performed measurement on the same rings but \emph{isolated}. We also observed an $h/e$ periodic signal, corresponding 
to persistent currents in the rings whose amplitude is similar to the 
one observed for connected rings. 

\begin{figure}[htbp]
    	\includegraphics[width=7.5cm]{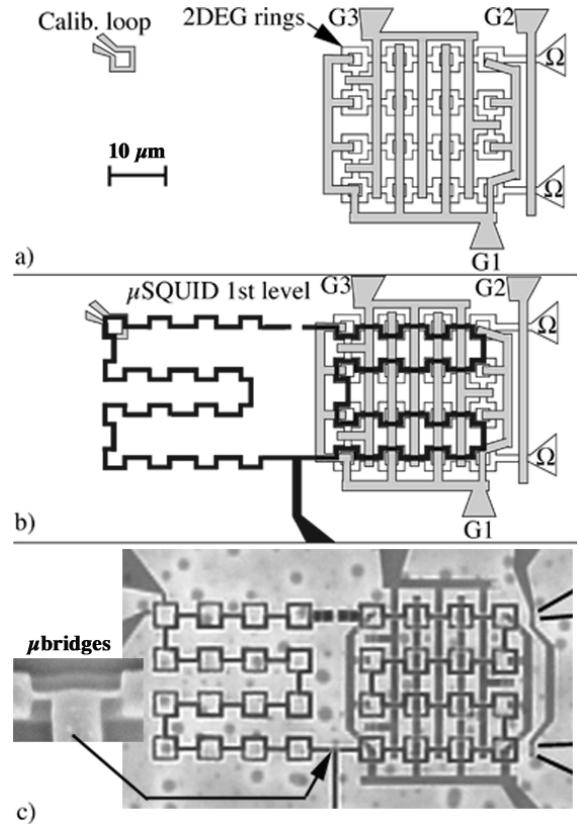}
	\caption{Fabrication of the sample: a) Etching of the rings and the 
	ohmic contacts (\textit{$\Omega$}), and deposition of the three 
	gates ($G_{1}$, $G_{2}$ and $G_{3}$) and the calibration loop ($calib.$ $loop$). b) First level 
	of the \MSB. c) Optical photograph of the sample, including the 
	second level of the \MSB. Inset shows an SEM picture of the 
	microbridges.}
	\label{figSample}
\end{figure}

The GaAs/GaAlAs heterojunction was grown using molecular beam epitaxy. 
The structure of the epilayers is $1\,\mu m$ GaAs 
buffer layer, $15\,nm$ GaAlAs spacer layer, $48\,nm$ Si doped GaAlAs 
layer and $5\,nm$ GaAs cap layer. At $4.2\, K$ in the dark, the two dimensional electron gas (2DEG) at the 
heterointerface had an electron density of $5.2\times 10^{11}\, cm^{-2}$ 
and a mobility of $0.8\times 10^{6}\, cm^{2}\, V^{-1}\, s^{-1}$.
This yields a Fermi velocity $v_{F}=3.16\times 10^{5}\, m\, s^{-1}$ and a Fermi 
wavelength $\lambda_{F} = 35\,nm$. All the lithographic operations were performed using electron beam 
lithography on PMMA (polymethyl methacrylate) resist with a 
JEOL 5DIIU electron beam writer. We first patterned an aluminium mask of 
the rings (see fig.\ref{figSample}), which was removed after etching $5\, 
nm$ of the GaAs cap layer by ion milling with $250\,V$ argon ions. This 
was sufficient to deplete the 
underlying 2DEG. The rings, actually squares of internal side length $2\,\mu 
m$, external side length $4\,\mu m$, mean perimeter $L=12\,\mu 
m$, connected by $2\,\mu m$ long arms, are connected to AuGeNi ohmic contacts 
in a two-probe geometry. Using a similar technique we fabricated 
wires of different widths to characterize the sample after etching. 
For wires of similar width, we measured a phase coherence 
length $l_{\phi}\approx 20\,\mu m$ derived from weak localisation. 
Conductance measurements gave an elastic mean free path of $l_{e} 
\approx 
8\,\mu m$, and a depletion length of $100\,nm$ 
on each side of the wire due to the etching process. Therefore the 
effective width of 
the arms of our rings is $W=0.8\,\mu m$. It should be noted that in 
this experiment, we 
were not in the pure diffusive case since $l_{e}\lesssim L$, but it is 
expected that the analytical results for the diffusive case still 
apply\cite{Mont2}. Three Schottky gates were then deposited allowing to deplete the 2DEG 
underneath. The first one ("$G_{1}$" in fig.\ref{figSample}) was deposited 
on top of each ring and allows all the interference effects 
in the rings
such as the persistent 
currents to be suppressed. 
The second gate ("$G_{2}$") was deposited on 
the two outgoing wires ("\textit{$\Omega$}") and made 
possible the insulation of the rings from the ohmic contacts.
The last one ("$G_{3}$") was placed 
on each arm joining two rings, allowing the measurement of isolated 
rings.
A calibration loop with the 
same 
dimensions as one ring was 
patterned in order to calibrate our experimental setup. 
Gates and calibration loop were obtained by liftoff of $10\,nm$ 
titanium and $50\,nm$ gold films. The device was then covered with a 
$60\,
nm$ insulating layer (AZ 1350 resist baked at $170^{\circ}C$).

The next step (see fig.\ref{figSample}) was the fabrication of the \MSB. It is designed as 
a gradiometer, equivalent to two counter-wound loops, 
in order to 
compensate the external magnetic flux\cite{Mailly,Cerni}, and made of 
aluminium because of its low critical current and the possibility to 
easily connect separate levels. The first 
level was deposited on top of the rings and of the calibration loop. Then a new 
insulating layer was 
deposited and the gradiometer was completed with the 
other half. The two 
Dayem micro-bridge Josephson junctions, $300\,nm\times 20\,nm\times 
30\,nm$, were lithographically defined at this 
stage.
The contact between the two aluminium levels was obtained by covering 
them with aluminium pads deposited after an ion bombardment cleaning. 
The first \MSA layer was made of $60\,nm$ thick aluminium and the second 
one, containing the two Dayem micro-bridges, of $30\,
nm$ thick aluminium in order to reduce the critical current of the \MSB. 
As the \MSA has exactly the same shape as the rings, the coupling 
between them is almost perfect\cite{Cerni}.

The gradiometer design and the accuracy of the e-beam lithography 
allow compensation of the external magnetic flux of $99.96\,\%$. The
magnetic field was swept from $-33.5\, G$ to $-3.5\, G$ at $7.5\, G\, s^{-1}$,
corresponding to $3\,\Phi_{0}$ and $9\,\Phi_{0}$ in each ring for the
inside and outside trajectories respectively: in this range of
magnetic field, the external magnetic flux is almost perfectly
compensated, and the gain of the gradiometer, calibrated by sending a
dc current through the calibration loop, nearly constant. Moreover,
such a limitation in the magnetic field sweep avoids flux penetration
in the Al layers of the \MSB. The critical current of the \MSA was
measured using dedicated electronics which ramps a dc current until
the \MSB's critical current is attained\cite{Cerni}.  The current is
then reset, and the value of the critical current sent to a computer. 
This cycle is repeated periodically at $5\, kHz$, and the measured
flux resolution of our gradiometer is then $\approx 5\times 10^{-4}\,
\Phi_{0}/ \!  \sqrt{Hz}$.

\begin{figure}[t]
	\includegraphics[width=7.8cm]{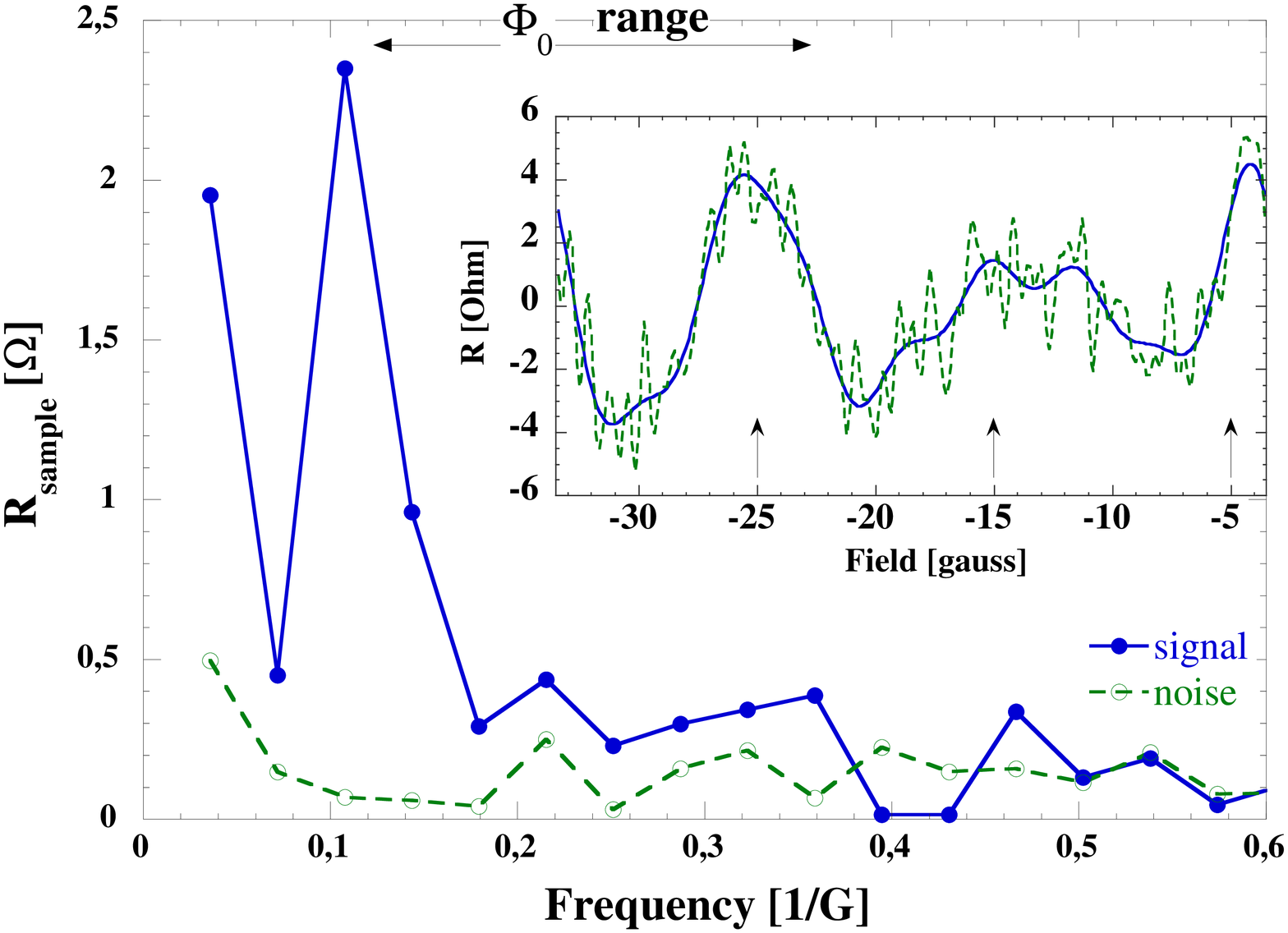}
	\caption{Typical power spectrum of the Aharonov-Bohm conductance 
	oscillations in a 
	line of 16 rings, measured at $20\, mK$. The horizontal arrow 
	indicates the $\Phi_{0}$ frequency 
	window, calculated from the geometrical parameters of our sample. The 
	Aharonov-Bohm peak is just one point out of this window, due to the 
	discretization on the frequency axis.
	Open circles 
	correspond to noise, i.e. FFT of the difference between 
	two succesive measurements. Solid circles correspond to signal. Increase of the signal at low frequency 
	is due to weak localisation and universal conductance fluctuations, 
	which are cancelled in the "noise" measurements. Inset shows the raw 
	data after substraction of the linear background (dashed line) and after 
	bandpassing the signal over the $\Phi_{0}$ frequency range (solid 
	line).}
	\label{figFFTAB}
\end{figure}

Resistance measurements as a function of magnetic field were performed separately using a conventional ac lock-in 
technique, with an ac current of $100\, pA$ at $777\, Hz$.
%The resistance 
%of the sample %($\approx 10\, k\Omega$ at $0$ field)
%is measured as a 
%function of the magnetic field which is swept from $-33.5\, G$ to $-3.5\, 
%G$. In this range of magnetic field, the external magnetic flux is 
%almost perfectly compensated, and the gain of the gradiometer, 
%calibrated by sending a dc current through the calibration loop, nearly 
%constant.
Fig.\ref{figFFTAB} shows the square root of 
the power spectrum of the resistance measured at $20\, 
mK$, obtained by Fast Fourier Transform (FFT) of the data. The horizontal scale shows the field frequency 
expressed in $G^{-1}$. Because of the aspect ratio of the rings, the $\Phi_{0}$ frequency extends from $0.12\,G^{-1}$ to 
$0.35\,G^{-1}$. The spectrum clearly exhibits a 
peak in the $\Phi_{0}$ frequency range, corresponding to Aharonov-Bohm 
conductance oscillations. The separation between each point of the FFT is directly 
related to the range of magnetic field used, which leads to a 
discretization of $0.036\, G^{-1}$ on the frequency axis.

\begin{figure}[t]
	\includegraphics[width=7.8cm]{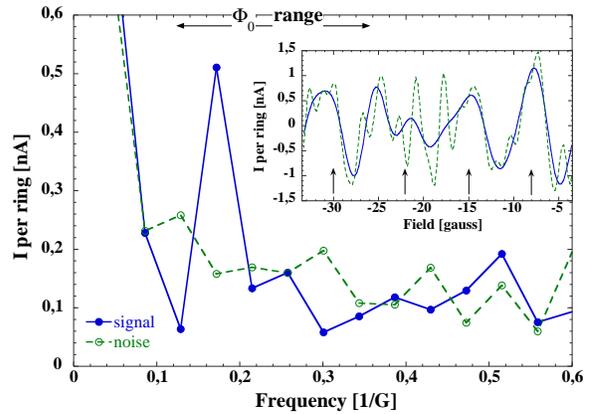}
	\caption{Typical power spectrum of the magnetization due to 
	persistent currents in a line of $16$ connected rings expressed in 
	$nA$ per ring, measured at 
	$20\, mK$. The horizontal arrow indicates the $\Phi_{0}$ frequency window, calculated 
	from the geometrical parameters of our sample. Open circles 
	correspond to noise, i.e. FFT of the difference between two 
	identical
	measurements. Solid circles correspond to signal, i.e. FFT of the difference between measurements with the 
	rings closed and opened. Inset shows the raw 
	data after substraction of the linar background (dashed line) and after 
	bandpassing the signal over the $\Phi_{0}$ frequency range (solid 
	line).}
	\label{figFFT}
\end{figure}

To measure the persistent currents, the critical 
current of the \MSA was measured as a function of the magnetic field. 
%again swept from $-33.5\, G$ to $-3.5\, G$ at $7.5\, G\, s^{-1}$. 
%The sensitivity to the currents in the rings is calibrated using the 
%calibration loop. 
We performed measurements with the rings "opened" and "closed"
($-561\,mV$ applied on the gate $G_{1}$ when measuring opened rings). 
The rings were insulated from the ohmic contacts by applying
$-558\,mV$ on the gate $G_{2}$.  Each measurement represented about
$1\, min$ of accumulation time.  The signal of persistent currents was
then extracted by FFT of the difference between measurements with
closed and opened rings, whereas the noise was obtained by FFT of the
difference between two identical measurements, either closed or opened
rings. Thus one spectrum (see fig.\ref{figFFT}) was obtained in about $4\, 
min$. This possibility of background noise subtraction is an
important advantage of our experimental technique\cite{Mailly}. The
total signal was divided by $16$ to obtain the current per ring.

A typical result for  
the square root of the power spectrum of the persistent currents expressed 
in $nA$ per ring is depicted in fig.\ref{figFFT}. The increase of the signal 
at low frequency is mostly due 
to slow temporal fluctuations of the critical current of the \MSB. Note that contrary to previous 
experiments\cite{Mailly}, the 
aspect ratio of the rings allows the signal to extend over several 
units of the FFT abscissa, corresponding to the $\Phi_{0}$ frequency 
range\cite{Washburn}: the amplitude of the persistent 
currents is then obtained by taking the difference between the area 
under the $\Phi_{0}$ region of the "signal" and the "noise" curves. By 
this means, the fact that we do not observe any increase of the 
amplitude of the "noise" curve in the $\Phi_{0}$ frequency range of 
fig.\ref{figFFT} but an increase of the amplitude of the "signal" curve
%the amplitude of the "signal" curve in the $\Phi_{0}$ 
%frequency range of fig.\ref{figFFT}
is a clear signature of persistent currents in our line of connected rings.

It should be stressed that random modifications of the disorder
configuration of the sample occured due to relaxation processes of the
dopants in the GaAlAs layer; this explains why the signal varied in
amplitude on a time scale of $\approx 15\, min$. It is this
instability that allows us to perform statistics over the amplitudes of the
persistent currents corresponding to
different disorder configurations, and thus to measure the typical
current\cite{Mailly}. Using this technique over a sample of $\approx
1000$ spectra, we find for our line of $16$ connected rings after 
substraction of the noise, a typical
current of $0.40\pm 0.08\,nA$ per ring.  This value has to be compared
with the theoretical value $0.58\times1.56\,(I_{0}/ \! 
\sqrt{N_{R}})\,(l_{e}/L)\approx 0.63\, nA$ per ring, $r=0.58$ being
the reduction factor predicted for the same line of connected rings
and $I_{0}$ being derived from our experimental parameters.

By applying $-540\, mV$ on the gate $G_{3}$ we isolated the rings and
performed the same experiment as for connected rings.  We also
observed an $h/e$ periodic signal, corresponding to persistent
currents in the rings of $0.35\pm 0.07\,nA$ per ring, to be compared
with the theoretical value $1.56\,(I_{0}/ \! 
\sqrt{N_{R}})\,(l_{e}/L)\approx 1.09\, nA$ per ring. Such a 
difference between experimental and theoretical values may be due to 
the finite value of
$l_{\phi}$\cite{Mont5} as well as the exact shape of the sample
(squares instead of rings).

%Note that strictly speaking the model used in 
%ref.\cite{Mont3} is valid only in the diffusive regime; in our 
%samples $l_{e}\lesssim L$, and deviations from the pure diffusive case may be 
%expected\cite{Mont4}. Finite value of $l_{\phi}$\cite{Mont5} as well as  the exact shape of the sample (squares instead of rings) may also account for the 
%difference between theoretical and experimental values. 
%Finally, contrary to recent 
%experiment on gold rings\cite{Mohanty}, the $h/2e$ component due to time 
%reversal paths is not 
%observed either in the Aharonov-Bohm conductance oscillations or in the persistent currents, 
%within the experimental uncertainty: this may be due to the fact that 
%such time reversal paths are less numerous in quasi-ballistic samples 
%than in diffusive ones.

%As mentioned 
%above, finite value of $l_{\phi}$ may account for the difference 
%+between experimental and theoretical values. 

The key point is the direct experimental comparison between persistent
currents in the same rings either connected or isolated.  Our
experiment gave a ratio $I_{connected}/I_{isolated}\approx 1.2\pm
0.5$. A similar experiment on a line of $4$ rings gave a similar
result.  The theoretical value calculated by ref.\cite{Mont3,Mont4}
for the diffusive case is $r\approx 0.58$: the difference may be due to
the fact that the model used is valid in the diffusive regime.  In our
samples $l_{e}\lesssim L$, and deviations from the pure diffusive case
may be expected\cite{Mont4}.  But such a ratio of order unity points
out the most striking result of our experiment: \emph{persistent
currents were not significantly modified when connecting or isolating
the rings}.  All previous experiments were carried out on
isolated rings\cite{Levy,Reulet,Mohanty,Chandra,Mailly}.  Our result
shows that persistent currents are not a specific property of isolated
systems: in our line of rings, electrons can visit the whole sample
and lose their phase coherence.  However, we have shown quantitatively
that the trajectories encircling individual loops, and thus giving
rise to the persistent currents, are very weakly affected by the
connection to arms.  This result should remain valid for any two
dimensional array of rings (or holes) whose perimeter is smaller than
$l_{\phi}$\cite{Mont3}.  By extension, this suggests that persistent
currents could be observed in a macroscopic sample: all the closed
trajectories smaller than $l_{\phi}$, which enclose flux, should give
rise to a measurable persistent current, even if the whole sample is
clearly not a quantum coherent object.

In conclusion, using a multiloop \MSA gradiometer, we have measured the magnetization of a line of 
connected GaAs/GaAlAs rings as a function of magnetic field. We have observed a 
periodic response, with period $h/e$; the amplitude of the 
corresponding persistent currents is in good agreement with 
theoretical estimates. Measurements on the same but isolated rings also 
showed an oscillatory 
component of the magnetization with period $h/e$: \emph{the amplitude 
of the persistent currents in connected and isolated rings was found to be similar}.

Further measurements on various geometries are now needed to fully 
understand the effect of the connectivity of the sample on the 
persistent currents. Experiments in the pure diffusive 
case or on a larger number of rings should 
also help to give a correct description of this "extensive" nature of 
persistent currents. From the 
theoretical side, a model for the ballistic case is needed for a 
direct comparison with our experiment.

Fruitful discussions with G. Montambaux, M. Pascaud, H. Bouchiat, V. 
Chandrasekhar and 
P. Butaud are acknowledged. We thank J. L. Bret, G. Simiand, J. F. 
Pini and H. Rodenas for technical help, D. K. Maude and J. 
Gilchrist for careful 
reading of the manuscript.

\end{document}